# AI in society and culture: decision making and values

**Katalin Feher**
Budapest Business School
University of Applied
Sciences, Hungary &
Drexel University
3141 Chestnut St,
Philadelphia, PA 19104,
USA
Fulbright Research Fellow
feher.katalin@uni-bge.hu

**Asta Zelenkauskaite**
Drexel University
3141 Chestnut St,
Philadelphia, PA 19104,
USA
az358@drexel.edu

## Abstract

With the increased expectation of artificial intelligence,
academic research face complex questions of human-
centred, responsible and trustworthy technology
embedded into society and culture. Several academic
debates, social consultations and impact studies are
available to reveal the key aspects of the changing
human-machine ecosystem. To contribute to these
studies, hundreds of related academic sources are
summarized below regarding AI-driven decisions and
valuable AI. In details, sociocultural filters, taxonomy of
human-machine decisions and perspectives of value-
based AI are in the focus of this literature review. For
better understanding, it is proposed to invite
stakeholders in the prepared large-scale survey about
the next generation AI that investigates issues that go
beyond the technology.

## CSS Concepts

Human-centered computing~Human computer
interaction (HCI) <concept_id>10003120</
concept_id>{Human-centered computing}

## Introduction

If digital technology becomes more complex, users will progressively empower the artificial intelligence (AI) to make decisions without human verification. In this change of decision making, users outsource certain activities to AI. The question is what kind of decision types are in the focus due to recent AI research and developments and what concepts orient the AI developments towards the human-centric approach?

Social science in its interdisciplinary context presents diverse landscape of AI-related research with constant human-centric and moral questions. The outline is focusing on this scope along human-machine decision making.

Our goal is to contribute to the CHI2020 workshop with a summary of the academic research landscape of AI-driven decisions and human-centric AI and provide future directions.

## Theoretical considerations

According to the inevitable trends, the AI related big data or big social data [1], human-machine augmentation or perceptive and fused intelligence [2], HCI design [3] or software-based narratives [4] are overloaded with choices and options with a wide selection. *"New forms of intelligence are making decisions in complex ways that escape the limits of human comprehension"* [5]. Therefore, users do not intend to control all online activity and to make all digital decision. Decisions and activities are getting more outsourced to smart or AI services considering only certain risks and benefits [6]. Transfer of control or empowered technology recall the constant dilemma of good AI society with the focus on responsibility [7] and trust (Europan Commission 2018, https://ec.europa.eu).

The next section presents an outline of current academic trends on this field.

## AI-driven decisions in sociocultural context

This overview is based on the literature review related to the issues of AI covered so far. First of all, the sociocultural context of AI is covered to understand the key subject areas. The number of related academic sources has been growing significantly in the recent years, therefore, hundreds of research papers have become available to analyze.

Our review is based on three fundamental academic databases, such as ArXiv as a repository of electronic preprints with strong focus on technological developments, Scopus as the largest abstract and citation database of peer-reviewed literature, and also, EbscoHost as a leading provider of research databases. After merging the outputs from the last five years as context of society and culture, and after the data cleaning, more than four hundred academic sources presented the database (n=432).

Two keywords were the absolute most frequent as the Excel tool in the studied academic abstracts, namely, "human" and "information". Searching the word pairs of these results, "structure of information" and "human being" were the most frequent. Based on the context of society and culture, these are assumed as filters of technologies to support or decline.

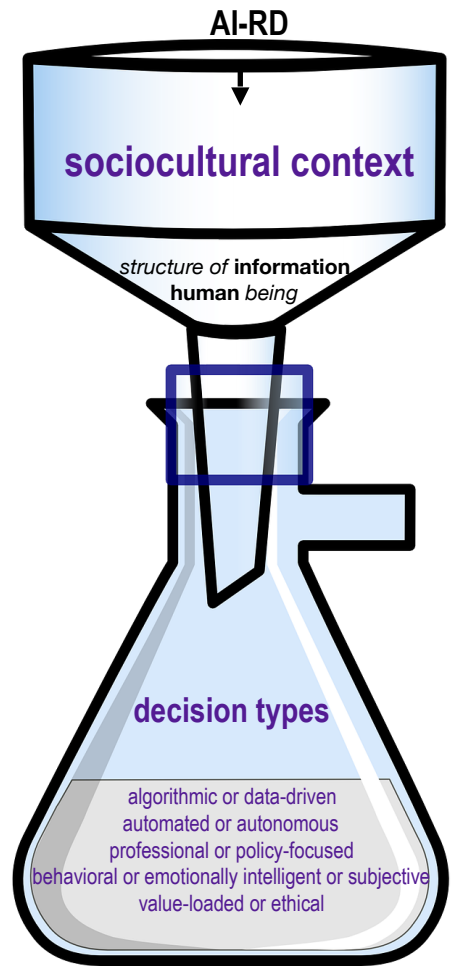

**Figure 1.** AI-driven decision types in sociocultural context according to academic publications (N=432)

Having these filters, the database was narrowed as topics of human-machine decisions. After data cleaning, almost twenty percent of the abstracts remained with relevant results. According to the findings applying manual content analysis [8], three categories of decision types have revealed as follows:

- AI-driven operation and decisions, such as algorithmic, data-driven, automated and autonomous decisions.

- Human-related decisions with options to be outsourced to the machines, serving the professional decision making or policy creation.

- Human decisions to preserve social-cultural values and to train trustable technology. Subjective, behavioral, emotionally intelligent, value-loaded or ethical decisions represent this category.

The boundaries are infrequently blurred between the categories, but the taxonomy was clearly drawn from the research topics. The laboratory funnel of Figure 1. represents a complex and slow process of the development of AI concepts where funnel filter works via currently available structure of information and human-driven perspectives extracted based on the recent research studies on a topic of AI. The filter passes through only those AI research & developments (AI-RD) which are adaptable as human-centric technologies. Decision types are critical to be filtered by academic research as control and power. The question that remains unanswered is how the trust is contributing to these contexts.

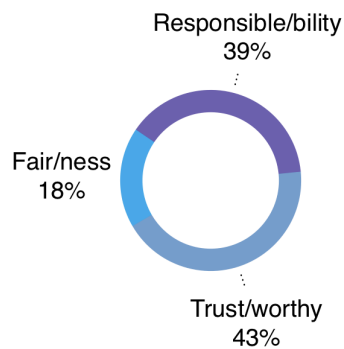

**Figure 2.**
Expected AI values
(N=48)

## Valuable AI

The transfer of control and the outsourced decisions presuppose a valuable technology to be trustable. One-tenth of the studied abstracts highlighted the importance of responsibility or trust or fairness. Trust and responsibility as values appeared almost equally on the top as critical requirement, while fairness was less represented (see Figure 2). According to the findings of the manual content analysis [8] of the reviewed studies, three approaches were revealed regarding valuable AI or good AI society:

- Issues of "responsibility" are moral and ethical questions to preserve the core tenets of humanity, geared to avoid discrimination or inequality. "Responsibility" is primarily mentioned along governmental or organizational policies.

- Topics of "trust" represent human-machine relationship with a strong focus on human-like robots, digital agents, and social-emotional intelligence. The focus is on the shared control or empowered AI.

- The subject of "fair" is less frequented. Fairness requires transparency and accountability for equality, creativity and respected diversity.

These values, expectations and requirements have become essential for academic research over the past five years. What are some implications of AI-values?

## Discussion and future direction

There are several implications of these findings. First, regardless of the number of the studies on AI and related fields, there is still a pressing need to further understand the AI-driven technologies. Second, definitions still focus heavily on values and trust issues that go beyond technological solutions. Those are rather conceptual, definitional, and broad questions of interest. As such, this means that application-based practices still are at the nascent phase, even if the field seems to be saturated with the AI-driven research.

Third, given the complexity in values, we would like to propose the next steps in this research to better understand what's the state of AI among various stakeholders. As such, future research should conduct targeted large-scale survey with various stakeholders who could provide insights for the next generation of AI research.

## Acknowledgement

Sincere gratitude to the Fulbright Commission for supporting this project.

## References

[1]  Asta Zelenkauskaite and Eric P. Bucy. 2016. A scholarly divide: Social media, Big Data, and unattainable scholarship. *First Monday,* 21, 5: https://doi.org/10.5210/fm.v21i5.6358

[2]  Yunhe Pan. 2016. Heading toward Artificial Intelligence 2.0. *Engineering*, 2, 4: 409-413. https://doi.org/10.1016/J.ENG.2016.04.018

[3]  Yang Chen. 2019. Exploring Design Guidelines of Using User-Centered Design in Gamification Development: A Delphi Study. *INT J HUM-COMPUT INT*, 35, 13: 1170-1181. https://doi.org/10.1080/10447318.2018.1514823

[4]  Simone Natale. 2018. If software is narrative: Joseph Weizenbaum, artificial intelligence and the

biographies of ELIZA. *New Media Soc*, 21, 3: 712-728. https://doi.org/10.1177/1461444818804980

[5]  Alexandre P. Casares. A. P. 2018. The brain of the future and the viability of democratic governance: The role of artificial intelligence, cognitive machines, and viable systems. *Futures*, 103, October: 5-16. https://doi.org/10.1016/j.futures.2018.05.002

[6]  Katalin Feher. 2019. Digital identity and online self: footprint strategies. An exploratory and comparative research study. *Int. J. Inf. Sci.* First Published: October 17. https://doi.org/10.1177/0165551519879702

[7]  Corinne Cath, Sandra Wachter, Brent Mittelstadt, Mariarosaria Taddeo and Luciano Floridi. 2018. Artificial Intelligence and the 'Good Society': the US, EU, and UK approach. *Sci. Eng. Ethics*, 24, 2: 505-528. https://doi.org/10.1007/s11948-017-9901-7

[8]  Klaus Krippendorf, K. 2004. Content Analysis. An Introduction to Its Methodology. London: Sage.